\documentclass[usenatbib]{mnras}

\usepackage{lmodern}

\usepackage[T1]{fontenc}
\usepackage[utf8]{inputenc}
\usepackage{color}
\usepackage{amsmath}
\usepackage{amssymb}
\usepackage{graphicx}
\usepackage[normalem]{ulem}
\usepackage{url}

\AtBeginDocument{\hypersetup{
linkcolor=linkblue,
citecolor=linkorange,
urlcolor=urlblue}}

\providecommand{\eprint}[1]{\href{http://arxiv.org/abs/#1}{#1}}
\providecommand{\adsurl}[1]{\href{#1}{ADS}}

\usepackage{ifthen}\def\eprinttmp@#1arXiv:#2 [#3]#4@{\ifthenelse{\equal{#3}{x}}{\href{http://arxiv.org/abs/#1}{#1}}{\href{http://arxiv.org/abs/#2}{arXiv:#2} [#3]}}
\renewcommand{\eprint}[1]{\eprinttmp@#1arXiv: [x]@}

\definecolor{urlblue}{rgb}{0,0,0.9}
\definecolor{linkblue}{rgb}{0,0,.8}
\definecolor{linkgreen}{rgb}{0,0.45,0}
\definecolor{linkpurple}{rgb}{0.7,0.0,0.4}
\definecolor{linkorange}{rgb}{0.7,0.1,0.0}

\makeatletter

\newcommand{\dd}{\textrm{d}}

\definecolor{somegreen}{cmyk}{0,0.49,0.98,0.09}
\definecolor{red}{rgb}{1,0,0}
\definecolor{magenta}{cmyk}{0,1,0,0}
\definecolor{violet}{cmyk}{0,1,0,0}
\definecolor{darkgreen}{rgb}{0,0.65,0.05}
\definecolor{antiquefuchsia}{rgb}{0.33, 0.1, 0.89}

\makeatother

\title[Measuring E(z) with standard candle clustering]{Measuring the Hubble function with standard candle clustering}

\author[Amendola \& Quartin]{Luca Amendola$^{1,\textcolor{blue}{\star}}$ and Miguel Quartin$^{2,3,\textcolor{blue}{\star} }$ \\
$^{1}$Institute of Theoretical Physics, Heidelberg University,
Philosophenweg 16, 69120 Heidelberg, Germany \\
$^{2}$Instituto de Física, Universidade Federal do Rio de Janeiro,
21941-972, Rio de Janeiro, RJ, Brazil\\
$^{3}$Observatório do Valongo, Universidade Federal do Rio de Janeiro,
20080-090, Rio de Janeiro, RJ, Brazil\\
$^{\textcolor{blue}{\star}}${\textsf{These authors contributed equally to this work.}}
}

\date{\today }

\begin{document}
\label{firstpage}
\pagerange{\pageref{firstpage}--\pageref{lastpage}}

\maketitle

\begin{abstract}
Supernova Ia magnitude surveys measure the dimensionless luminosity distance $H_{0}D_{L}$. However, from the distances alone one cannot obtain quantities like $H(z)$ or the dark energy equation of state, unless further cosmological assumptions are imposed. Here we show that by measuring the power spectrum of density contrast and of peculiar velocities of supernovae one can estimate also $H(z)/H_{0}$ regardless of background or linearly perturbed cosmology and of galaxy-matter bias. This method, dubbed Clustering of Standard Candles (CSC) also yields the redshift distortion parameter $\beta(k,z)$ and the biased matter power spectrum in a model-independent way. We forecast that an optimistic (pessimistic) LSST may be able to constrain $H(z)/H_{0}$ to 5--13\% (9--40\%) in  redshift bins of $\Delta z=0.1$ up to at least $z=0.6$.
\end{abstract}


\maketitle

\section{Introduction}

The measurement of the cosmic expansion rate $H(z)$ as a function of redshift $z$ is one of the central tasks of observational cosmology. An accurate knowledge of $H(z)$ allows to directly constrain the cosmological model and, with further hypotheses, to extract the main background parameters such as the equation of state of dark energy and the abundance of dark matter. Several techniques have been employed so far to measure $H(z)$, each one with pros and cons.

The most important method exploits the apparent magnitude of standard candles, e.g.~supernova Ia (SNe), to get $E(z)\equiv H/H_{0}$ \citep{Perlmutter:1998np,riess1998}. However, standard candles measure the dimensionless luminosity distance $H_{0}D_{L}(z)$, not $E(z)$, and the two quantities are one-to-one related in a FLRW universe only if one knows the spatial curvature. In isotropic but inhomogeneous models, e.g.~LTB models, one should also know the curvature at every redshift~\citep{2006PhRvD..73h3519A}.

Transverse and longitudinal baryon acoustic oscillations (BAO), on the other hand, give both $E(z)$ and the dimensionless angular-diameter distance $H_{0}D$~\citep{2003ApJ...598..720S}. However, this relies on the assumption that the BAO peaks are not displaced by the galaxy bias  or by the growth function (which in principle are  space-dependent), or by intrinsic distortions due to a non-standard inflationary process. Moreover, the non-linear effects, that also affect the BAO scales, must be modelled accurately~\citep{Crocce:2007dt,Padmanabhan:2009yr,Alam:2016hwk}.

Cosmic chronometers \citep{Jimenez_2002} are also often employed. Here many systematic uncertainties about the passive evolution of stars must be under control to a high level~\cite[see e.g.][]{Liu_2015}. Ultimately, this can be done only relying on population synthesis simulations based on standard physics and cosmology. Any form of modified gravity, to give an example, would make such assumptions very uncertain.

Gravitational wave standard sirens allow to determine $D_L(z)$ but, again, not directly $H(z)$~\citep{Schutz:1986gp}. Time delays of sources at $D_s$ due to strong gravitational lenses at $D_d$ allow the determination of $D_d D_s/[D_{ds} (1+\kappa_{\rm ext})]$, where $\kappa_{\rm ext}$ is the convergence on the line of sight~\cite[see e.g.][]{2007ApJ...660....1O}. Time delays are thus also incapable of directly delivering $H(z)$. Moreover, the method requires a detailed modelling of both the lens and line of sight (for $\kappa_{\rm ext}$), which can at the moment be achieved only for a few systems~\citep{Wong:2019kwg}. Further combining time delays \citep{Denissenya:2018zcv,Collett:2019hrr} or strong lensing \citep{Rasanen:2014mca} with supernovae distances, one can in principle get the spatial curvature and, in a FLRW model (but not in a more general isotropic model), $E(z)$.

One can measure $H(z)$ directly through the redshift drift effect~\citep{Sandage1962} without assuming homogeneity~\citep{Uzan:2008qp,Quartin:2009xr}. This however requires next generation spectrographs, ample observing time and very large telescopes~\citep{Liske:2008ph}.

In this paper, we show that the linear power spectra (of both density and velocities) of SNe can provide another method to measure $E(z)$ which is free of the limitations mentioned above. In particular, we show that this technique, dubbed Clustering of Standards Candles (CSC), is independent of assumptions concerning cosmology (both background and linear perturbations) or the bias. The CSC method could be applied in principle to any type of standard candles, from galaxies obeying the Tully-Fisher~\citep{TullyFisher1977,Masters:2007ud} or Fundamental Plane relation~\citep{DjorgovskiDavis1987,Springob:2014qja} to  high-redshift quasars~\citep{2019NatAs...3..272R} and standard sirens. For the latter, the CSC will directly deliver $H(z)$.

A combined analysis of these linear power spectra for the case of galaxies was investigated by~\cite{Burkey:2003rk}.  The use of SNe in this context was first proposed by~\cite{gordon2007}. In~\cite{Castro:2015rrx} it was shown that for SNe the velocity power spectrum and weak-lensing observables are very complementary, and an uncertainty of $0.2$ on $\sigma_8$ was inferred with current SNe. \cite{Odderskov:2016qig}  used $N$-body simulations to investigate the importance of   realistic spatial distributions in the SNe mock catalogs. The combination of all three spectra here considered was used to measure the growth-rate by~\cite{Adams:2017val} and forecasts for LSST were made by~\cite{Howlett:2017asw} and later by~\cite{Graziani:2020kkr}. The observability of the velocity power spectrum was analyzed in detail in~\cite{Garcia:2019ita} for many observational strategies. All these works, however, assumed a specific cosmology or parametrization. More importantly, they did not realise that these observables could be combined to measure $E(z)$. In~\cite{2006PhRvL..96s1302B} a different method, based on the SNe luminosity distance dipole, has been proposed to determine $E(z)$, but it requires the knowledge of the spatial curvature. In~\cite{Huterer:2016uyq,Boruah:2019icj} SNe velocities were used to measure $f \sigma_8$ at low redshifts where the dependence on cosmology is weak. Finally, in~\cite{Mukherjee:2018ebj} a generalization of the Alcock-Paczynski test employing a combination of SN and galaxies was proposed which can further constrain the background cosmology and does not rely on linear perturbation theory.

Of course, the CSC technique does rely on other hypotheses, and is therefore to be considered complementary to the methods currently pursued. In particular, we assume that SNe are well-standardized candles, that the Etherington relation between luminosity and angular-diameter distance is satisfied, that we are well into the linear regime, and that matter obeys the continuity equation. However, these hypotheses do not concern the cosmological model we are  investigating, and can be tested independently through other methods.

\section{Model-independent variables}

Let us start from the linear continuity equation at sub-horizon scales $\partial\delta_m/\partial t = -(1+z)\nabla \cdot \boldsymbol{v}$, where $\delta_m$ is the matter density contrast and $\boldsymbol{v}$ the peculiar velocity (henceforth PV) vector field.
In  Fourier space, if a matter tracer density contrast $\delta_{T}$  is given by $\delta_{T}=b\delta_m$, where $b$ is the bias, and $f \equiv d\log\delta_m/d\log a$ is the growth rate,   the continuity equation becomes
\begin{align}
    \boldsymbol{v } =i H\beta \frac{\boldsymbol{k}}{k^{2}(1+z)}\,\delta_{T},
\end{align}
where $\beta=f/b$. However, what we measure in general is only the longitudinal velocity $v_{\parallel}=\boldsymbol{v}\cdot\boldsymbol{r}/r$~\citep[although see][]{Hotinli:2018yyc}, so the relation
becomes
\begin{equation}
    v_{\parallel}=i\frac{H}{k(1+z)}\beta\frac{\boldsymbol{k}\cdot\boldsymbol{r}}{kr}\delta_{T}
    =i\frac{H\mu}{k(1+z)}\beta\delta_{T}\,,\label{eq:cont}
\end{equation}
where $\mu=\cos\theta_{\boldsymbol{k},\boldsymbol{r}}$ is the angle between $\boldsymbol{k}$ and the line of sight $\boldsymbol{r}$. We can independently measure $\beta(k,z)$ from the power spectrum redshift distortion~\citep{Guzzo2008}. We thus see that measuring both peculiar velocities and density fluctuations, we can estimate the combination $H\mu/k$. However, in order to measure $\mu,k$ from the raw data (angular separations and redshifts), one needs a cosmological model. Since we aim at being model independent, we need to find out how $\mu,k$ transform with the cosmological model.

Measuring $v_{\parallel}$ requires an independent estimation of distance. One can measure the PV of a standard candle by  assuming that sources with the same apparent magnitude are at the same distance. In this case in fact  the residual difference in the observed redshift must be due to linear peculiar velocities,  correlated with the density fluctuations, and  to uncorrelated components, namely calibration/experimental errors, and the non-linear component of the PV. This is the path we follow below.

Let us now consider two Gaussian fields with zero mean: $\delta_T$ and $v_{\parallel}$, sampled by a single population of SN with number density $n_{\rm SN}$, and let $\delta,v$ represent their $k$-th Fourier coefficients. The SNe are expected to faithfully trace galaxies, although we do not need this assumption.

In the linear regime, the only relevant correlations are the density-density, the velocity-density and the velocity-velocity correlations~\citep{Burkey:2003rk,Howlett:2017asw}.
Let $P_{m}\equiv V\langle\delta_{m}\delta_{m}^{*}\rangle$ be the linear matter power spectrum  and  $P\equiv b^{2}n_{\rm SN}P_{m}$  the signal-to-noise SN spectrum. We then get
\begin{align}
    & \!\!\!\! n_{\rm SN}V\langle\delta\delta^{*}\rangle  =n_{\rm SN}\big(1+\beta\mu^{2}\big)^{2}\,S_{\delta}^{2}b^{2}\,P_{m} +1 = B^{2}\,PN_{\delta},\!\!\!\!\!\label{Pdd}\\
    & \!\!\!\! in_{\rm SN}V\!\langle\delta v^{*}\rangle  \!=\! \frac{n_{\rm SN}H\mu}{k(1+z)}\big(1+\beta\mu^{2}\big)S_{\delta}S_{v}\,\beta b^{2}P_{m} \!=\! ABP \,, \!  \label{Pdv}\\
    & \!\!\!\! n_{\rm SN}V\!\langle vv^{*}\rangle \!=\! n_{\rm SN}\!\left[\frac{H\mu \, S_{v}}{k(1+z)}\right]^{2}\!\!\! \beta^{2}b^{2} P_{m} \!+\! \sigma_{v,{\rm eff}}^{2} \!=\! A^{2}PN_{v}. \!\!\label{Pvv}
\end{align}
Although not much is known of the SN bias $b$, as will be shown our method is insensitive to this quantity. The non-linear smoothing factors $S_{\delta},S_{v}$, important only at small scales, are taken to be~\citep{2014MNRAS.445.4267K,Howlett:2017asw}
\begin{align}
    S_{v}  = \frac{\sin (k\sigma_{v})}{k\sigma_{v}}\,, \quad
    S_{\delta} = \frac{1}{\sqrt{1+0.5(k\mu\sigma_{\delta})^{2}}} \, ,
\end{align}
where $\sigma_{v},\sigma_{\delta}$ are  assumed for the moment to be constant.  Note that while $\sigma_\delta$ is the pairwise source velocity dispersion, $\sigma_v$ is just a phenomenological parameter first introduced in~\cite{2014MNRAS.445.4267K}. We also defined
\begin{align}
    A &\equiv \frac{H\mu\beta S_{v}}{k(1+z)} \, ,\\
    B &\equiv (1+\beta\mu^{2})S_{\delta} \, ,
\end{align}
and $N_{\delta}=1+1/(B^{2}P)$, $N_{v}=1+\sigma_{v,{\rm eff}}^{2}/(A^{2}P),$ are the shot-noise factors. Here the power spectrum $P$ includes the growth function and depends arbitrarily on $k,z$.

Any statistical uncertainty $\sigma_{\rm int}$ in the magnitude of a standard candle is associated via the distance modulus relation to an uncertainty in the redshift, which generates a  uncertainty in velocity given by~\citep{2006PhRvD..73l3526H,Davis:2010jq}:
\begin{equation}
    \sigma_{v,{\rm eff}}^{2}\equiv\left[\frac{\log10}{5}\sigma_{\rm int}\right]^2\!\left[2-\frac{d\log D_L}{d\log(1+z)}\right]^{-2}\!\!\!+\frac{\sigma_{v{\rm ,nonlin}}^{2}}{c^{2}}.\label{eq:new-sv-1}
\end{equation}
We assume a constant  $\sigma_{\mathrm{int}}=0.13$~mag (but we explore other values later on) and $\sigma_{v{\rm ,nonlin}}=300$ km/s; the impact of $\sigma_{v{\rm ,nonlin}}$ is always subdominant. Lensing introduces a $z-$dependent scatter which in $\Lambda$CDM is $\sim 0.05z$~mag~\citep{Jonsson:2010wx,Quartin:2013moa}, which can be neglected here since in our redshift range it adds little to our assumed value for $\sigma_{\rm int}$. Systematic errors in distances would also affect the zero-point of velocities, but this was found to be  negligible in~\cite{Howlett:2017asw}.

Note that the second square bracket in Eq.~\eqref{eq:new-sv-1} embodies the model-independent relation (assumed for the moment to be perfectly measured) between the observed  magnitude scatter $\delta m$ neglecting $\sigma_{v{\rm ,nonlin}}$) and the inferred  $v_{\parallel}$. In a flat FLRW space~\citep{2006PhRvD..73l3526H}
\begin{equation}
    \delta m = \frac{v}{c}\left[1-\frac{(1+z)^2}{HD_L}\right]
    \approx -(1+z)^2 \,\frac{v}{c} \frac{1}{HD_L}
\end{equation}
for $z\ll 1$. Then, writing the correlations in terms of the observable $\delta m$, rather than $v_{\parallel}$, the $H$ dependence would cancel out; this cancellation however breaks down at finite~$z$.

As  mentioned, however, without a cosmological model, we cannot derive $\mu,k$ from observations, since we cannot convert the raw observables into distances. Rather, we are forced to use an arbitrary reference cosmology, e.g.~$\Lambda$CDM, to convert angles and redshifts into $\mu_{r},k_{r}$. This induces several modifications to the formalism above. If we choose arbitrarily a reference cosmology (subscript $r$)  $\mu$ depends on the true cosmological model
as $\mu=\mu_{r}H/(H_{r}\alpha)$ and $k$ as $k=\alpha k_{r}$, where ~\citep{2000ApJ...528...30M,Amendola:2004be}
\begin{equation}
    \alpha \,=\, \frac{H}{H_{r}} \frac{\sqrt{\mu_{r}^{2}(\eta^{2}-1)+1}}{\eta}
\end{equation}
and
\begin{equation}
    \eta \,\equiv\, \frac{HD}{H_{r}D_{r}} \,.
\end{equation}
Then we have
\begin{equation}
    \frac{H\mu}{k} \,=\, \frac{H_{r}\mu_{r}}{k_{r}} \frac{ \eta^{2}}{\big[\mu_{r}^{2}(\eta^{2}-1)+1\big]} \,.
    \label{eq:expr-eta}
\end{equation}
Since we use $\log\eta=\log HD-\log H_{r}D_{r}$ as variable (thus producing automatically relative errors) the reference model does not affect the final Fisher matrix (FM) entry for $\log\eta$. Also, since we integrate over $\mu_{r}$ between $[-1,+1]$ (naturally the same range of $\mu$), the reference model has no impact on the $\mu$ integral as well.

Eq.~\eqref{eq:expr-eta} shows that what really enters $\langle\delta v^{*}\rangle$ and $\langle v v^{*}\rangle$ is the combination $HD$, rather than $H$ alone. We conclude that we should replace everywhere $A,B$ with $\bar{A},\bar{B},$ where
\begin{align}
    \bar{A} & =\frac{H_{r}S_{v}}{k_{r}(1+z)}\frac{\beta\eta^{2}\mu_{r}}{\mu_{r}^{2}(\eta^{2}-1)+1}  \,, \\
    \bar{B} & =\left[1+\frac{\beta\eta^{2}\mu_{r}^{2}}{\mu_{r}^{2}(\eta^{2}-1)+1}\right]S_{\delta} \,.
\end{align}
Moreover, the argument of $P$ is also rescaled
\begin{equation}
    P(k)\,=\,P(k_{r}\alpha) \,=\,P\!\left(\frac{k_{r}H}{H_{r}}\frac{\sqrt{\mu_{r}^{2}(\eta^{2}-1)+1}}{\eta}
    \right).
\end{equation}
At this point we need to assume we know $H_{0}D$ from SNe magnitude
surveys. Then from $\eta$ and $H_{0}D$ we can replace $H$ with $\eta H_{r}D_{r}/D$
 and we only need to evaluate
\begin{equation} \label{eq:Pcorr}
    \frac{\partial P}{\partial\eta} \,=\, P'k_{r} \frac{D_{r}}{D}\, \frac{\eta\mu_{r}^{2}}{\sqrt{\mu_{r}^{2}(\eta^{2}-1)+1}},
\end{equation}
where $P'=dP/dk$. The same considerations also apply to $\beta$.
The dependence on the volume in $\bar n$ and in $P_m$ cancels out, so no further correction for $P$ is needed. The same rescaling of $k,\mu$ occur in  $S_{\delta,v}$, which then depend only on $\eta,D$ and on the reference values $D_r,k_r,\mu_r$.

To perform forecasts we need to choose a fiducial cosmology. This, however, is not a break of our model-independent approach: it is just due to the fact that we do not yet possess real data; when they will be available, the fiducial model will be replaced by the measured data. Moreover, since we take as fiducial $\Lambda$CDM with both bias and growth-rate $f$ independent of scale, we have $\beta'=0$ and the $\beta$ correction analogous to Eq.~\eqref{eq:Pcorr} has no effect on our forecasts, although of course it has to be included to perform forecasts for  more general cases.

Since we do not want to parametrize 
$P(k,z)$ or $\beta(k,z)$, we need to split the observations into $k,z$-cells. So we take $P,\beta$ as free parameters in each cell. Since the $k$-cells are independent in the linear regime we can simply sum the FMs over them. We conclude that the CSC method applied to SN surveys allows the reconstruction of $H_{0}D(z), \,E(z)$, $P(k,z)$, $\beta(k,z)$,  $\sigma_{\delta}$ and $\sigma_{v}$ without any parametrization or choice of the cosmological model. With several tracers, beside increasing the precision, one can also measure $\beta_{i},\sigma_{v_{i}},\sigma_{\delta_{i}}$ for each species $i$~\citep[see e.g.][]{Seljak:2009,McDonald:2009,Abramo:2013,Abramo:2019ejj}. Quantities like $f,b$ and $P_{m}$ remain inaccessible without assumptions on the cosmological model or the bias function.

Given $E(z)$ and $d=H_{0}D(1+z)$, one can directly measure the present spatial curvature as   $\Omega_{k0}=[(Ed_{,z})^{2}-1]/d^{2} $. While in any FLRW model this quantity should be independent of $z$, it will in general depend on $z$ in isotropic but inhomogeneous Universes. Within $\Lambda$CDM, although $\Omega_{k0}$ has been measured to be close to zero to high precision in the CMB~\citep[see][]{Aghanim:2018eyx}, some subsets of the data favor non-zero curvature~\citep{DiValentino:2019qzk}.

\section{Fisher matrix for model-independent variables}

As we have seen, the random variables $x_{a}=\sqrt{n_{\rm SN}V}\{\delta,v\}$ are  Gaussian with zero mean and covariance matrix
\begin{equation}
    C_{ab}=P\left(\begin{array}{cc}
    \bar{B}^{2}N_{\delta} & \bar{A}\bar{B}\\
    \bar{A}\bar{B} & \bar{A}^{2}N_{v}
    \end{array}\right) .
\end{equation}
As in \cite{Abramo:2019ejj} we write the FM  for a set of parameters $\theta_{\alpha}$, in a survey of volume $V$, as~\citep[see also][]{Tegmark:1997rp}
\begin{equation}
    F_{\alpha\beta} \,=\, \frac{1}{(2 \pi)^3} 2\pi k^{2}\Delta_{k}V\bar{F}_{\alpha\beta} \,=\, VV_{k}\bar{F}_{\alpha\beta}\,,
\end{equation}
where $V_{k}= (2\pi)^{-3}2\pi k^{2}\Delta_{k}$ is the volume of the Fourier space after integrating over the azimuthal angle but not over the polar angle (i.e., the volume of a spherical Fourier space shell of width $\Delta_{k}$ would be given by $\int \dd \mu V_k$). In this expression $\bar{F}$ is the FM per unit phase-space volume $V V_k$ integrated over $\mu$,
\begin{equation}
    \bar{F}_{\alpha\beta}=\frac{1}{2}\int_{-1}^{+1}d\mu\frac{\partial C_{ab}}{\partial\theta_{\alpha}}C_{ad}^{-1}\frac{\partial C_{cd}}{\partial\theta_{\beta}}C_{bc}^{-1} \,,
\end{equation}
where the integrand is evaluated at the fiducial value. If we fix $\sigma_\delta$ and  $\sigma_v$ we are left with $\theta_{\alpha}=\{\log P,\log\beta,\log\eta\}$, for which  we obtain a 3$\times$3 FM for each $k,z$ cell. From now on we assume $H_{0}D$ is known with high accuracy from SNe magnitudes, and therefore a constraint on $\eta$ is entirely equivalent to a constraint on $E(z)=H/H_{0}$.

\begin{table}
    \footnotesize
    \setlength{\tabcolsep}{2.2pt}
    \begin{tabular}{ ccc | cc | cc}
    & & & \multicolumn{2}{c}{\rm{LSST}~$20\%$} & \multicolumn{2}{|c}{\rm{LSST~SQ}} \\
    $z_{\rm bin}$ & V & $k_{\rm min}$ & $10^{3}\cdot n_{\rm SN}$ &  $\Delta H/H$ & $10^{3}\cdot n_{\rm SN}$ & $\Delta H/H$ \\
      & (\text{Gpc}/$h)^{3}$ & (h/\text{Mpc}) &(h/\text{Mpc}$)^3$  & (\%) & (h/\text{Mpc}$)^3$ & (\%) \\
    \hline
     0.05 & 0.046 & 0.0175& 0.064 & 13.2 & 0.011  & 39  \\
     0.15 & 0.296 & 0.0094& 0.07  & 8.9  & 0.032  & 14   \\
     0.25 & 0.727 & 0.0070& 0.076 & 7.6  & 0.054  & 9.2  \\
     0.35 & 1.27  & 0.0058& 0.081 & 6.9  & 0.051  & 8.6  \\
     0.45 & 1.88  & 0.0051& 0.087 & 6.3  & 0.037  & 9.9  \\
     0.55 & 2.51  & 0.0046& 0.093 & 5.8  & 0.012  & 20   \\
     0.65 & 3.13  & 0.0043& 0.099 & 5.4  & 0.0019 & 100   \\
     0.75 & 3.72  & 0.0041& 0.10  & 5.1  & 0.0002 & -
    \end{tabular}
    \caption{\label{tab:sn-surveys}  Survey specifications and corresponding forecasts for a 5-yr LSST survey (18000 deg$^{2}$) for both cases here considered. The $z$ bins have $\Delta z=0.1$ and are centred on $z_{\rm bin}$.
    For LSST 20\% we assume 20\% completeness of SNe in all redshifts, while LSST SQ (Status Quo) is based on the current provisional strategy, which yield much lower completeness.
    }
\end{table}

To obtain the actual errors for a given $(k,z)$-cell we multiply the specific FM by $V V_k$. For a $z$-shell of volume $V(z)$ and for
$\Delta_{k}\approx2\pi/V^{1/3}$,
\begin{equation}
   VV_{k} = \frac{k^{2}V^{2/3}}{2\pi}\, .
\end{equation}
While $\beta,P$ depend on $k,z$, the parameter $\eta$  depends only on $z$. We can thus combine observations in different $k$-bins of the same $z$-shell to obtain the overall constraints on $\eta(z)$ and thus on $E(z)$. The $k$-cells are chosen with equal
$\Delta_{k}=2\pi/V(z)^{1/3}$ between $k_{\rm min}(z)$ and $k_{\rm max}$. Following~\cite{Garcia:2019ita}, $k_{\rm min}=2\pi/V(z)^{1/3}$
(see Table~\ref{tab:sn-surveys}),  while $k_{\rm max} = 0.1~h/$Mpc ensures we remain in the linear regime. Since our method does not require a parametrization of $P(k)$, it can be employed also in the mildly non-linear regime, provided the redshift distortion factor $(1+\beta\mu^2)$ in Eqs.~\eqref{Pdd}--\eqref{Pdv} and the continuity equation~(\ref{eq:cont}) are still valid approximations. We thus test also up to $k_{\rm max}=0.3 ~h/$Mpc as in the LSST science requirements~\citep{Mandelbaum:2018ouv}. For the $i$-th $k$-cell we have a matrix whose only non-zero elements are
\begin{equation}
    F_{i}=VV_{k}
    \left(\begin{array}{cccc}
    ... & ... & ... & ...\\
    ... & B_{k_{i}} & ... & B_{k_{i}z}\\
    ... & ... & ... & ...\\
    ... & B^T_{k_{i}z} & ... & B_{z}
    \end{array}\right) \,,
\end{equation}
where $B_{k_{i}}$ is the 2$\times$2 block of the $k$-dependent variables $(P,\beta$) for the $i$-th $k$-bin, $B_{z}$ is the $k$-independent quantity~$\eta$, and $B_{k_{i}z}$ the mixed entries. We then sum over the $F_{i}$ for all the $k$-cells obtaining a large  $[2\cdot n_k+1]^2$ matrix, where $n_k\approx 0.1/\Delta_k$ is the number of $k$-cells in a given $z$ bin (between 5 and 24 for our $\Delta_k$). We finally invert this matrix, and extract the final joint errors on $P(k),\beta(k),\eta$ for that $z$-shell as the square root of the corresponding diagonal entry. Note that this procedure is not equal to just summing the 3$\times$3 FMs independently for each $k$-cell since $\eta$ is correlated with all the $k$-dependent quantities.

\begin{figure}
    \includegraphics[width=.99\columnwidth]{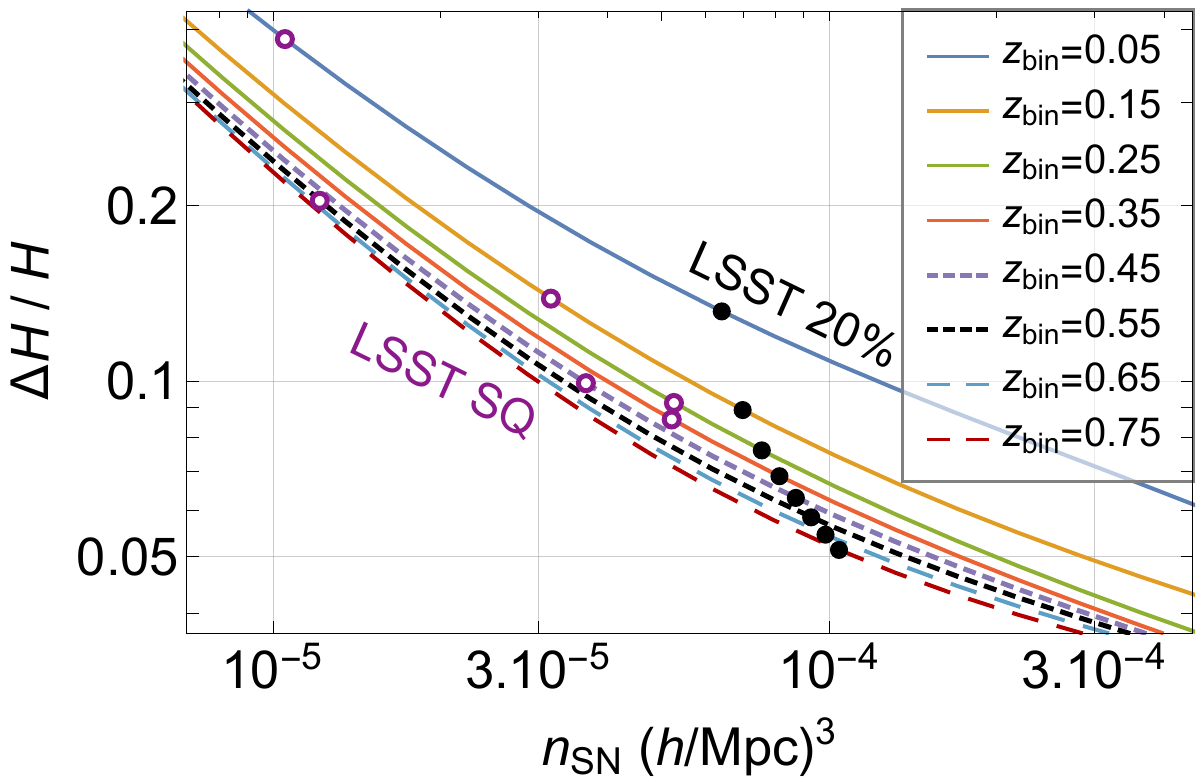}
    \caption{\label{fig:precision-vs-nsn} Marginalized relative errors for $H(z)$ as a function of the number density of SNe in the various $z$ bins for the two LSST surveys of Table~\ref{tab:sn-surveys}. The estimates for LSST 20\% (LSST SQ) are depicted by solid black (open purple) dots. Note that LSST is far from the cosmic variance limit: larger $n_{\rm SN}$ will lead to better precision.
    }
\end{figure}

\section{Forecasts}
For our   numerical estimates, we assume as fiducial values $\eta=1$, $P_{m}$ from $\Lambda$CDM with Planck 2018 values~\citep{Aghanim:2018eyx} and non-linear Halofit corrections as implemented in CAMB~\citep{LewisChallinorLasenby2000,Takahashi:2012em}, $b=1.6$~\citep{Howlett:2017asw,Mukherjee:2018ebj}, $\beta=\Omega_{m}^{\gamma}/b$ with $\gamma=0.545$, $\sigma_{d}=4.24$ Mpc$/h$ and $\sigma_{v}=13$ Mpc$/h$~\citep{2014MNRAS.445.4267K,Howlett:2017asw}. Following~\cite{Garcia:2019ita}, we consider both an optimistic and a pessimistic case for the LSST survey, as detailed in Table~\ref{tab:sn-surveys}. We dub the former LSST 20\%, since it assumes a simple constant 20\% SN completeness for $z<0.8$, and the latter LSST SQ for Status Quo of the public LSST observational strategy, including all photometric quality cuts used in~\cite{Abell:2009aa}. We also assume a SN rate which is a good fit to~\cite{Cappellaro:2015}: $2.1 \cdot 10^{-5} (1+z)^{1.95}$/(yr Mpc${}^3$). We find that $\Delta H/H$ can be constrained in the LSST 20\% survey from 13\% at low $z$s to 5\% at high $z$s (Table \ref{tab:sn-surveys}). For the SQ survey the sharp decrease of the high-$z$ SN density induces a fast weakening of the constraints for $z > 0.5$. Figure~\ref{fig:precision-vs-nsn} illustrates how the forecasts depend on the number densities of SNe in different $z$ bins.  Finally we remark that our forecasts vary by less than $3\%$ when changing the fiducial by $\pm 2\sigma$ from Planck 2018 values.

\begin{figure}
    \includegraphics[width=.99\columnwidth]{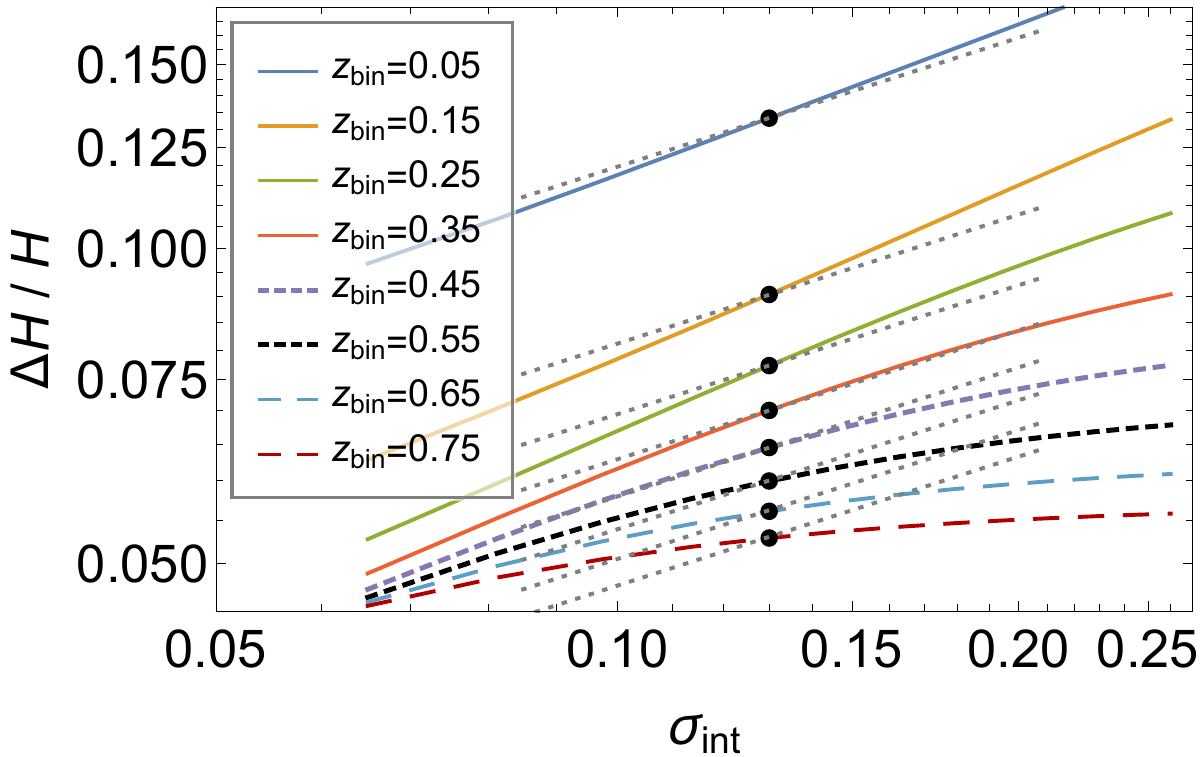}
    \caption{\label{fig:precision-vs-sint} Marginalized relative errors for $H(z)$ as a function of the intrinsic scatter $\sigma_{\rm int}$ in the various $z$ bins for the LSST 20\% survey. The dashed lines depict the power-law approximation of Eq.~\eqref{eq:finfit}, and the black dots the fiducial value of $\sigma_{\rm int}$.  Note that dependence is stronger at low-$z$ and weaker at higher $z$s.
    }
\end{figure}

As expected, the error depends mainly on the number density $n_{\rm SN}$ of supernovae, on the redshift, and on the absolute magnitude uncertainty $\sigma_{\mathrm{int}}$. An approximate formula which is typically valid within 9\% in our range of $z$ and $n_{\rm SN}$, and for $\sigma_{\mathrm{int}}\in(0.065,0.26)$, is
\begin{equation}
    \frac{\Delta H}{H}\approx 0.064\left(\frac{\sigma_{\rm int}}{0.13}\right)^{0.41} \left(\frac{z}{0.3}\right)^{-0.29}\left(\frac{n_{\rm SN}}{10^{-4}}\frac{\rm{Mpc}^3}{h^3}\right)^{-0.54} \! .
    \label{eq:finfit}
\end{equation}
As can be seen from Figure~\ref{fig:precision-vs-nsn} and Eq.~\eqref{eq:finfit} the method works far from the cosmic variance limit. In particular for LSST-like surveys the errors decrease roughly as $n_{\rm SN}^{-1/2}$, which is halfway between the shot-noise limit ($n_{\rm SN} P_m \ll 1$ for $\langle \delta \delta^* \rangle$, $n_{\rm SN} P_m \ll [\sigma_{v,{\rm eff}} \,k/H]^2$ for $\langle v v^{*} \rangle$) in which $\Delta H\propto n_{\rm SN}^{-1}$ and the cosmic variance limit ($n_{\rm SN} P_m \gg 1$, $[\sigma_{v,{\rm eff}} \, k/H]^2$) in which $\Delta H\propto n_{\rm SN}^{0}$. The dependence on $\sigma_{\rm int}$ is illustrated in more detail in Figure~\ref{fig:precision-vs-sint}, where it can be seen that it is a bit stronger at lower redshifts (the exponent is closer to 0.5 for $z\le 0.4$) and weaker at higher ones.

We can also relate the relative uncertainties in $H$ with the ones in $D_L$. The latter scales roughly as $\sigma_{\rm int}/\sqrt{N(z)}$, so for larger $\sigma_{\rm int}$ the relative uncertainties in $D_L$ degrade faster than the ones in $H$. To wit, $\Delta D_L/D_L \propto (\Delta H/H) \, \sigma_{\rm int}^{0.59}$. If we again follow the LSST science requirements~\citep{Mandelbaum:2018ouv} and assume a degradation of the distance uncertainties by $22\%$ due to systematics we get
\begin{equation}
    \frac{\Delta H}{H} \,\approx\, 400 \left(\frac{z}{0.3}\right)^{0.56} \left(\frac{\sigma_{\rm int}}{0.13}\right)^{-0.59}  \frac{\Delta D_L}{D_L} .
    \label{eq:H-vs-D-fits}
\end{equation}
where $\Delta D_L/D_L$ is the average uncertainty of the luminosity distance in a given redshift bin centered at $z$ by combining all SN in that bin. This means that if standard candles can give a 0.01\% distance uncertainty at $z=0.3$ with traditional methods, they would also give a $\sim 4\%$ measurement of $H(z)$. Also, it confirms that in computing errors in $H$ from $\eta$ one can neglect the much smaller errors in $D_L$.

We assumed so far that the relation between magnitude scatter and velocities, $\delta m=g(z) v/c$, where
\begin{equation}
    g(z) \,\equiv\, \frac{5}{\log 10} \left[2-\frac{\dd\log D_L}{\dd\log(1+z)}\right]
\end{equation}
was precisely known. This requires a broad assumption that the theoretical
$D_L(z)$ be a reasonably smooth function so one can measure it in many $z-$bins and interpolate. If one allows features in $D_L(z)$ with abrupt changes in arbitrarily small $\Delta z$ the CSC cannot be employed as proposed. Following the reasoning in the previous paragraph, one can estimate the errors in $g(z)$ from the error on $[D_L(z+\Delta z) - D_L(z)]$. For LSST $20\%$ using bins of $\Delta z = 0.01$ ($0.03$) one gets a relative error of $<10\%$ ($<2\%$) for $g(z)$, which validates our assumptions in practice.

The fitting formula~\eqref{eq:finfit} also illustrates that the CSC method does not have a strong dependence on $\sigma_{\rm int}$, and in particular weaker than the one found in~\cite{Garcia:2019ita} for measuring $\sigma_8$ using only  $\langle v v^\star\rangle$, because $\langle\delta\delta^\star\rangle$ and $\langle\delta v^\star\rangle$ do not depend on this parameter. Therefore, even if the Hubble diagram scatter for LSST ends up being larger than 0.13~mag, for instance due to the lack of spectra of most SNe, the method will remain competitive. In fact, a relative increase in $\sigma_{\rm int}$ is more than offset by a similar relative increase in $n_{\rm SN}$.

For LSST 20\%, increasing $k_{\rm max}$ to 0.2 $h/$Mpc reduces uncertainties significantly and the relative errors in $H$ become \{11\%, 7.1\%, 5.6\%, 4.6\%, 3.9\%, 3.4\%, 3.1\%, 2.8\%\} for $\,z=0.05\,$ up to $\,z=0.75$. In other words, they shrink by 20\% at $z=0.05$ and by 40\% for $z\ge 0.4$. Using $k_{\rm max} = 0.3$ $h/$Mpc yields less significant improvements, with uncertainties decreasing by an additional $4\%$ (at $z=0.05$) to $18$\% (for $z\ge0.4$). A doubling (halving) of $P'$ improves (worsen) $\Delta H/H$ by 25\%. The dependence on $\sigma_{\delta,v}$ is weak: leaving both free with weak priors leaves results almost unchanged.

\begin{figure}
    \includegraphics[width=.97\columnwidth]{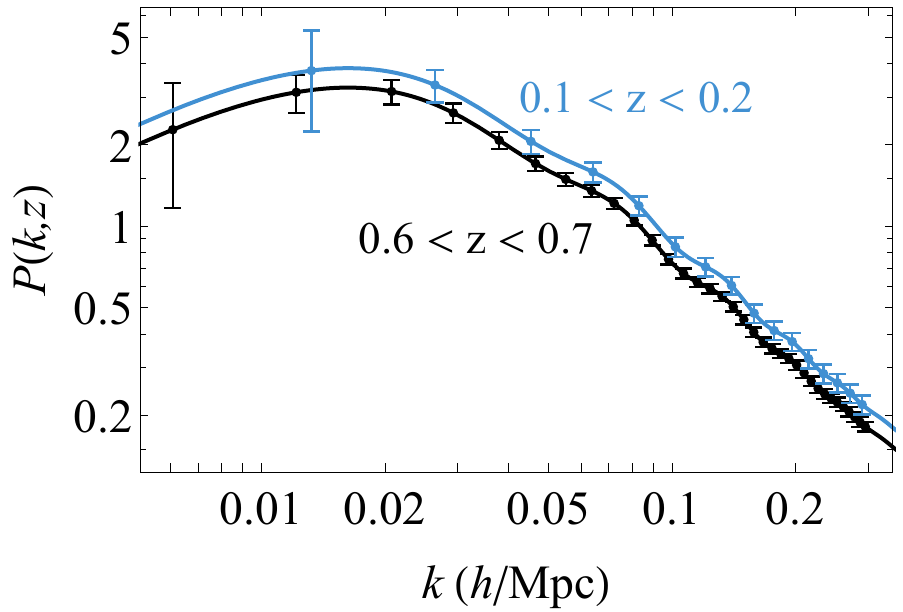}
    \caption{\label{fig:errors-Pk}  $1\sigma$ errors on the supernova power spectrum $P(k,z)$ for two redshift bins for the optimistic LSST 20\% SN catalog and assuming $\Delta k = 4\pi/V(z)^{1/3}$, which is twice as large as used in the text in order to have less cluttering in the plot.}
\end{figure}

Two examples of the error on $P(k,z)$ are depicted in Figure~\ref{fig:errors-Pk} for LSST 20\%. A  complete exploration of the uncertainty landscape, with more accurate fits and  with multiple tracers, will be provided in a future publication.

\section{Conclusions}
In the next decade the number of observed SNe will increase by orders of magnitudes.  We have shown that it is possible to estimate $H(z)/H_0$ using SN data only and
without restricting to any particular cosmological model and without assumptions on either the galaxy-matter bias or the SN bias. This method, which we dubbed the clustering of standard candles, or CSC, can be extended to any class of sources with a standardized observable, i.e.~calibrated to depend only on~$z$.

Although the parametrization of cosmological quantities like $H(z),P(k)$ etc.~allows one to get stringent constraints because of the projection of large volumes of raw phase-space data onto a small number of parameters, it is in some cases possible to obtain important information projecting (i.e., binning) only relatively small phase-space cells. This information is therefore by construction independent of parametrization and can be applied to any cosmological model. This might help avoid  confirmation biases inherent to the choice of only a very small subset of possible models, for instance $\Lambda$CDM and its variants.
The trade-off between model-independence and statistical uncertainty is worth being explored in full.

\section{Acknowledgements}
We would like to thank Bruno Moraes, Arman Shafieloo and Shinji Tsujikawa for interesting discussions, many of which held with LA  during the 6th Korea-Japan meeting on Dark Energy at KMI Nagoya. MQ is supported by the Brazilian research agencies CNPq and FAPERJ. Both Mathematica notebooks which can be used to reproduce the FM  calculations can be downloaded at \url{www.github.com/itpamendola/clustering} and \url{www.github.com/mquartin/clustering}.

\section{Data availability}
The data underlying this article will be shared on reasonable request to the corresponding author.

\bibliography{references,scaling_bib}

\label{lastpage}
\end{document}